\documentstyle[11pt,newpasp,twoside,epsf]{article}
\def\edcomment#1{\iffalse\marginpar{\raggedright\sl#1\/}\else\relax\fi}
\reversemarginpar

\begin{document}
\title{Detection of negative superhumps in a LMXRB -- an end to the 
long debate on the nature of V1405~Aql (X1916-053)} 

\author{A.~Retter}
\affil{School of Physics, University of Sydney, 2006, Australia;
retter@Physics.usyd.edu.au; and Dept. of Physics, Keele University, 
Keele, Staffordshire, ST5 5BG}
\author{Y. Chou}
\affil{Department of Physics, National Tsing Hua University, No. 101, 
Sec. 2, Kuang Fu Rd, Hsinchu, Taiwan; yichou@phys.nthu.edu.tw}
\author{T. Bedding}
\affil{School of Physics, University of Sydney, 2006, Australia;
bedding@Physics.usyd.edu.au}

\begin{abstract}

Two similar periodicities (3001 and 3028~s) are known from the X-ray 
and optical light curves of V1405~Aql, a low mass X-ray Binary 
(LMXRB). Two competing models have been offered for this system. 
According to the first, V1405~Aql is a triple system. The second model 
invokes the presence of an accretion disc that precesses in the apsidal 
plane, suggesting that the shorter period is the orbital period while 
the longer is a positive superhump. Re-examination of previously published 
X-ray data on V1405~Aql reveals an additional periodicity of 2979~s. The 
periods in V1405~Aql fit well within a newly found relation where the ratio 
between the negative superhump deficit (over the orbital period) and the 
positive superhump excess is a function of orbital period in cataclysmic 
variables that show both types of superhumps. Therefore, the 2979-s period 
is naturally interpreted as a negative superhump. The recently found 4.8-d 
period in the X-ray light curve of V1405~Aql is consequently understood 
as the precession of the accretion disc in the nodal direction. This is 
the first firm detection of negative superhumps and nodal precession in a 
LMXRB. Our results thus confirm the classification of V1405~Aql as a 
permanent superhump system. The 13-year argument on the nature of this 
intriguing object has thus finally come to an end. 

\end{abstract}

\section{Introduction}

Permanent superhump systems show superhumps (quasi-periodicities shifted 
by a few percent from their orbital periods) in their optical light 
curves. The phenomenon is observed during their normal brightness state,
unlike in SU~UMa systems. Permanent superhumps can either be a few percent 
longer than the orbital periods and are then called `$\bf positive$ 
$\bf superhumps$', or shorter than the orbital periods -- `$\bf negative$ 
$\bf superhumps$'. The positive superhump is explained as the beat 
between the binary motion and the precession of an accretion disc in the 
apsidal plane. Similarly, the negative superhump is understood as the 
beat between the orbital period and the nodal precession of the disc 
(Patterson 1999).

Permanent superhumps have been observed in about 20 cataclysmic variables. 
Superhumps have been seen in a few LMXRBs in outburst (e.g. O'Donoghue \& 
Charles 1996), however, there have not been any confirmed detection of 
permanent superhumps in a LMXRB.

White \& Swank (1982) and Walter et al. (1982) independently found 3001-s
periodic dips in the X-ray light curve of the LMXRB V1405~Aql. Schmidtke 
(1988) and Grindlay et al. (1988) reported a detection of an optical 
periodicity at 3028~s. The difference between the X-ray and optical 
periods was confirmed by further extensive observations of V1405~Aql.

Two basic models have been offered so far for the periodicities found in 
V1405~Aql. According to the first model (Grindlay et al. 1988), the longer 
3028-s period is the binary inner orbital period, while the shorter 3001-s 
period is the beat period between the binary period and the $\sim$4-d 
orbital period of a third companion. The second model (White 1989) suggests
that the 3001-s period is the binary period and that the 3028-s period is a 
positive superhump. The debate on the nature on V1405~Aql has still been 
continued (Chou, Grindlay, \& Bloser 2001; Homer et al. 2001; Haswell et 
al. 2001). Here we argue that the second model is correct, and present 
evidence for a third period that we identify as the negative superhump. For
more details see Retter, Chou \& Bedding (2001).

\section{Observations and Analysis}

We have re-analysed existing RXTE data that were presented by Chou et al. 
(2001). In Fig.~1a the power spectrum of 10 successive runs in 1996 May 
is presented. In addition to the two known periods (3001 and 3028~s, 
marked as f$_{1}$ and f$_{2}$) and their 1-d aliases, there is a third 
peak (f$_{3}$) together with its 1-d alias pattern. After fitting and 
subtracting the two known frequencies, the third, which corresponds to 
the periodicity 2979.3$\pm$1.1 s, becomes the strongest peak in the 
residual power spectrum (Fig.~1b). To test the possibility that a 
combination of the window function and the two known periods is 
responsible for the third periodicity, we planted sinusoids of the two 
known periods in the data (plus noise). The result (Fig.~1c) did not 
show any other significant peak in the power spectrum. We also rejected 
manually the points corresponding to the dips from the light curve. The 
new period was still present in the corresponding power spectrum (Fig.~1d). 
Finally, we divided the data into two parts (first and last five runs), 
and the new periodicity appeared in both.

\section{Discussion}

The 2979-s period is shorter than the 3001-s period by about 0.7\%. 
Assuming that the 3001-s period is the orbital period and that the 
3028-s period is a positive superhump, the new period is naturally 
explained as a negative superhump. Patterson (1999) proposed that the 
negative superhump deficit is about half the positive superhump excess. 
The corresponding ratio in V1405~Aql (0.8) is somewhat larger than this. 
In Fig.~2 we show this ratio for all systems known to have both positive 
and negative superhumps, and we see a clear trend as a function of orbital 
period. Our result for V1405~Aql fits this trend very well. 

\begin{figure}

\plottwo{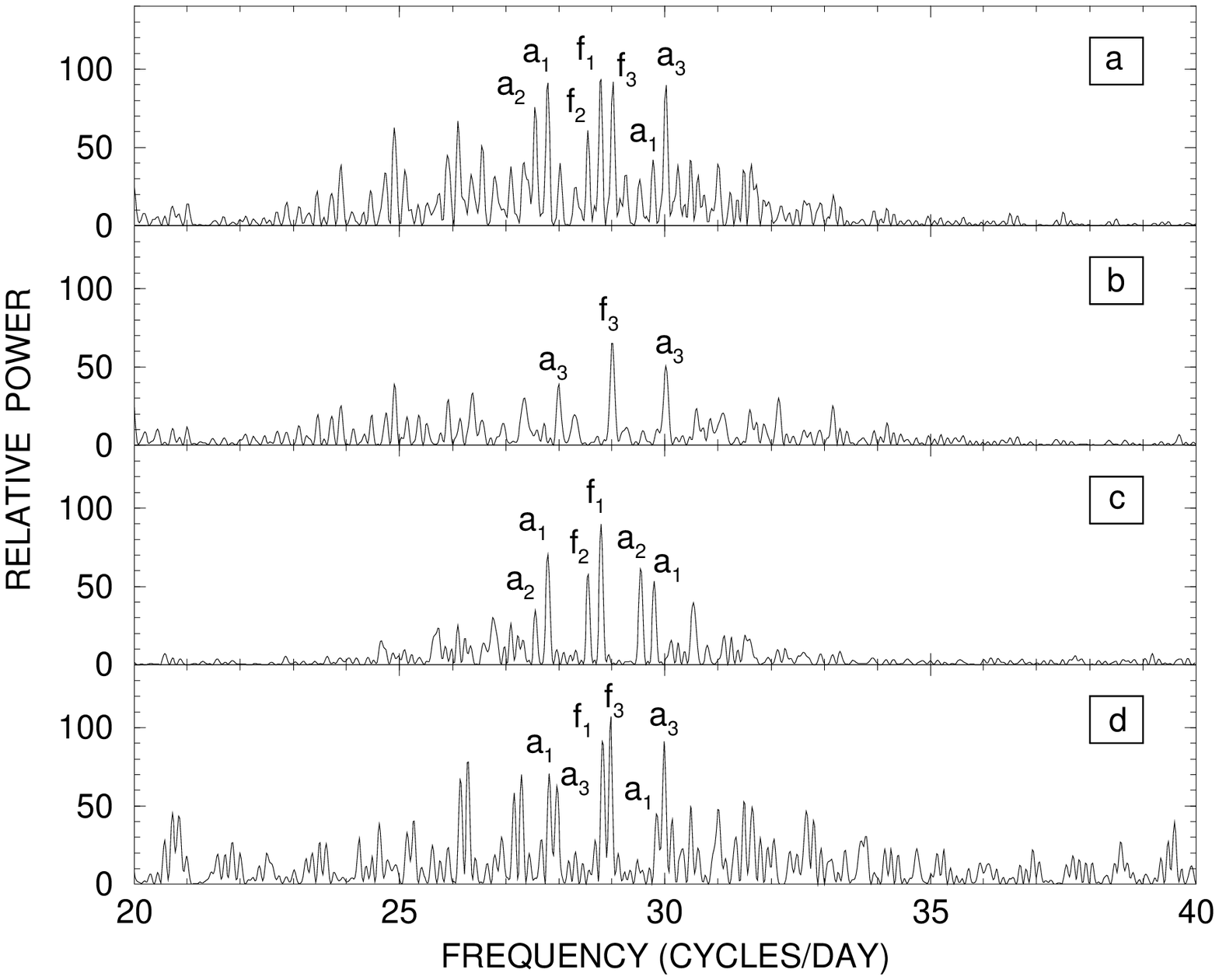}{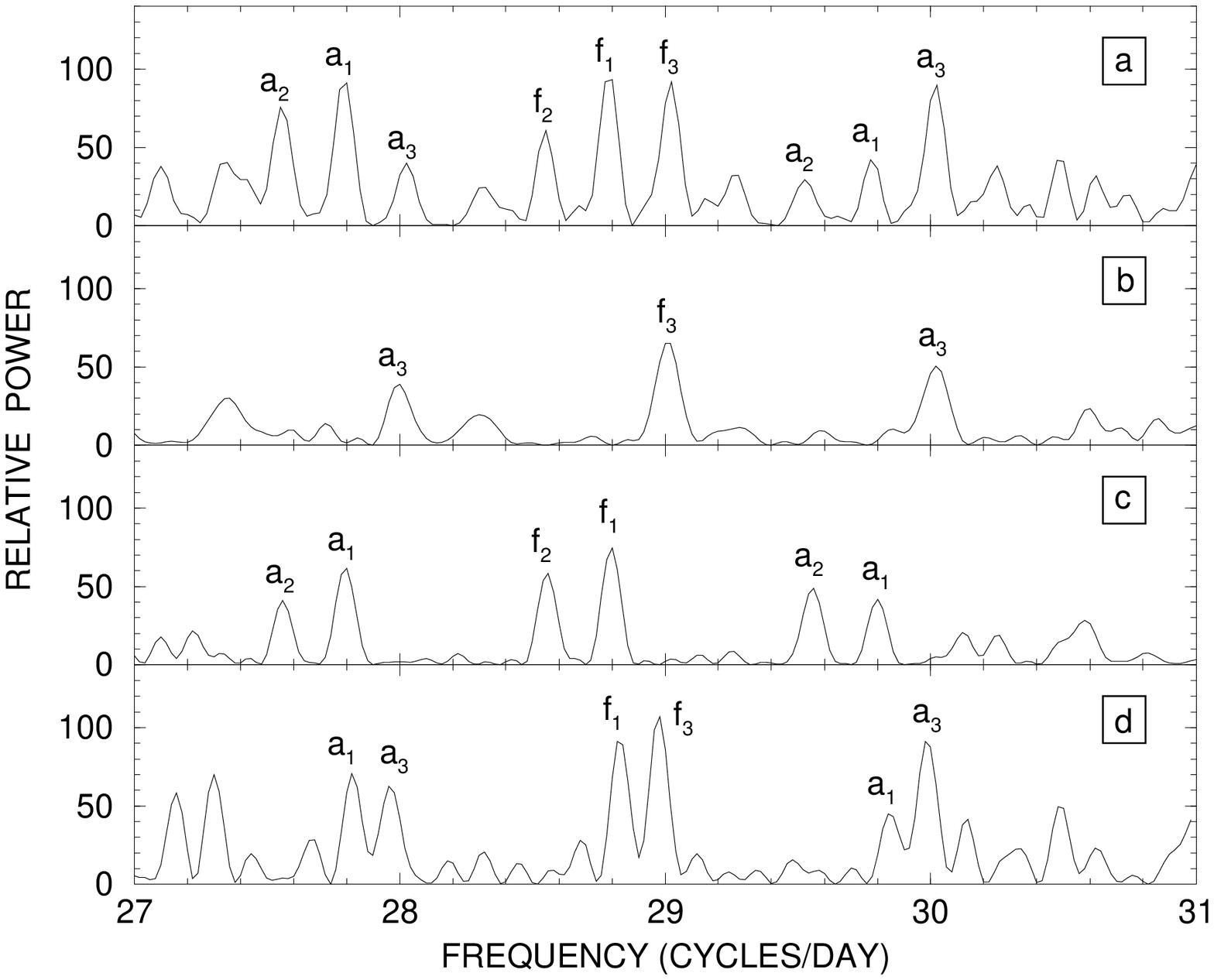}

\caption{Power spectra of 10 successive runs in 1996 May: 
{\bf a}. Raw data; In addition to the two previously known periods -- the 
3001-s period (marked as f$_{1}$) and the 3028-s period (f$_{2}$), there is 
a third structure of peaks centered around 2979~s (f$_{3}$); `a$_{i}$' 
(i=1-3) represent 1-d aliases of `f$_{i}$'.
{\bf b}. After fitting and subtracting f$_{1}$ and f$_{2}$, f$_{3}$ is 
still present and becomes the strongest peak in the power spectrum.
{\bf c}. Power spectrum of a synthetic light curve, consisting of 
sinusoids of the two previously known periods (plus noise) sampled as the 
data thus illustrating the window function. This shows that aliases of the 
known periods cannot explain the f$_{3}$ peak.
{\bf d}. Same as (a) after rejecting the dips. The peak at f$_{3}$ 
dominates the power spectrum. This test confirms that this frequency is 
not a consequence of random variations in the structure of the dips.}

\end{figure}

Our suggestion that the 2979-s period is a negative superhump implies a 
nodal precession of $\sim$4.8 d. Indeed Chou et al. (2001) found that 
the phase jitter of the X-ray dips in the 1996 May data is modulated 
with a period of 4.86 d. Homer et al. (2001) reached a similar conclusion 
from a different dataset and found a period of 4.74$\pm$0.05 d. Therefore, 
the superhump model can explain this periodicity as well. The classification 
of V1405~Aql as a permanent superhump system is thus firmly established, 
and our result puts an end to the 13-year debate on the nature of this 
intriguing system.

\begin{figure}

\plotone{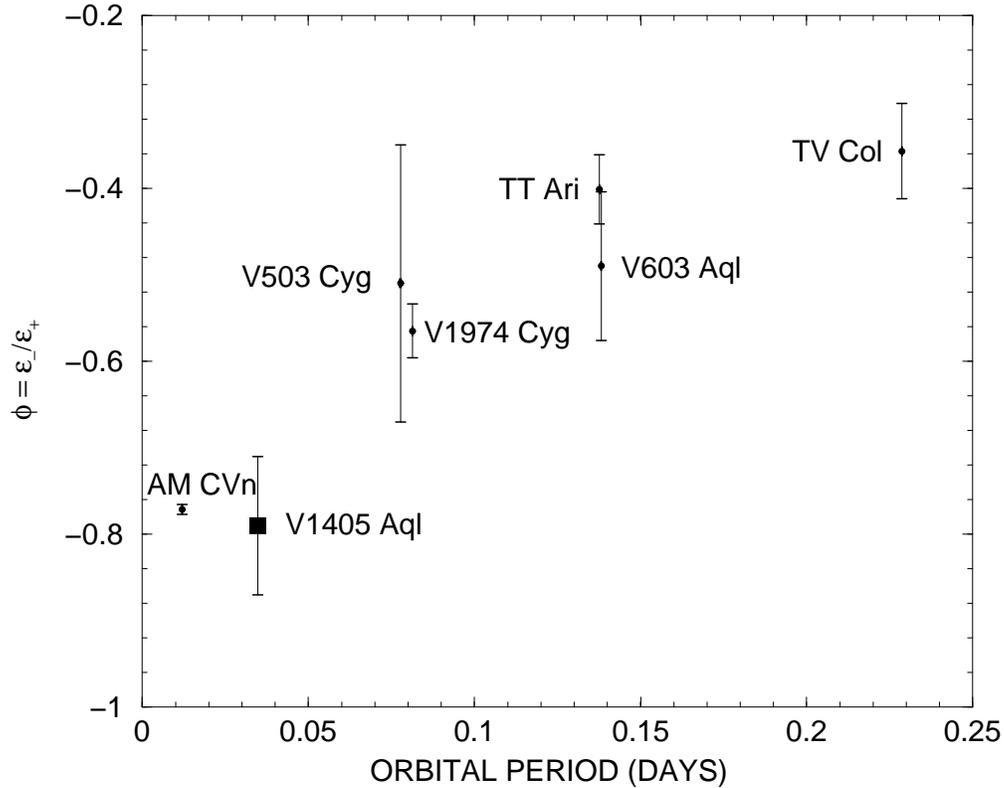}

\caption{The relation between the orbital period and the ratio between the
negative superhump deficit and the positive superhump excess in systems
that have both types of superhumps. The periods in V1405~Aql obey this 
relation.}

\end{figure}

\end{document}